\documentclass[acmsmall, screen]{acmart}

\AtBeginDocument{%
  \providecommand\BibTeX{{%
    \normalfont B\kern-0.5em{\scshape i\kern-0.25em b}\kern-0.8em\TeX}}}

\usepackage{calligra}
\usepackage{algorithm}
\usepackage{algorithmicx}
\usepackage{algpseudocode}

\usepackage{tabularx}
\usepackage{array}
\usepackage{xspace}
\usepackage{booktabs}
\usepackage{moreverb}
\usepackage{fontenc}
\usepackage{amsmath}
\usepackage{fancybox}
\usepackage{color}
\usepackage{colortbl}
\usepackage{array}
\usepackage{multirow}
\usepackage{multicol}
\usepackage{listings}
\usepackage{makecell}
\usepackage{graphicx}
\usepackage{setspace}
\usepackage[multiple]{footmisc}
\usepackage{wasysym}

\usepackage{threeparttable}

\usepackage{balance}
\usepackage{csvsimple}
\usepackage{longtable}
\usepackage{caption}
\usepackage{url}
\usepackage{lscape}
\usepackage{rotating}
\usepackage{tikz}
\usepackage{enumitem}
\usepackage{subcaption}

\usepackage{cuted} 
\usepackage{stfloats} 

\usepackage{marvosym}
\usepackage{pifont}
\usepackage{courier}
\usepackage{afterpage,natbib,lipsum}

\usepackage[utf8]{inputenc}
\usepackage[T1]{fontenc}
\usepackage{titlesec, lipsum}

\titleformat{\paragraph}[block]{\em\filcenter}{}{0pt}{}
\titleformat{\availability}[block]{\em\filcenter}{}{0pt}{}

\algnewcommand\algorithmicforeach{\textbf{for each}}
\algdef{S}[FOR]{ForEach}[1]{\algorithmicforeach\ #1\ \algorithmicdo}

\newcolumntype{L}[1]{>{\raggedright\let\newline\\\arraybackslash\hspace{0pt}}m{#1}}
\newcolumntype{C}[1]{>{\centering\let\newline\\\arraybackslash\hspace{0pt}}m{#1}}
\newcolumntype{R}[1]{>{\raggedleft\let\newline\\\arraybackslash\hspace{0pt}}m{#1}}

\definecolor{codegreen}{rgb}{0,0.6,0}
\definecolor{codered}{rgb}{1,0,0}
\definecolor{codegray}{rgb}{0.5,0.5,0.5}
\definecolor{codepurple}{rgb}{0.58,0,0.82}
\definecolor{backcolour}{rgb}{0.95,0.95,0.92}
\definecolor{lightgray}{gray}{0.9}

\lstdefinestyle{mystyle}{
    commentstyle=\color{codegreen},
    keywordstyle=\color{magenta},
    numberstyle=\small\color{black},
    stringstyle=\color{codepurple},
    basicstyle=\scriptsize\ttfamily,
    breakatwhitespace=false,
    breaklines=true,
    captionpos=b,
    keepspaces=true,
    showspaces=false,
    showstringspaces=false,
    showtabs=false,
    tabsize=2
}

\lstdefinelanguage{diff}{
  morecomment=[f][\color{blue}]{@@},     
  morecomment=[f][\color{red}]-,         
  morecomment=[f][\color{codegreen}]+,       
  morecomment=[f][\color{red}]{---}, 
  morecomment=[f][\color{codegreen}]{+++}
}

\lstdefinelanguage{text}{
  breaklines=false
}

\setlist{noitemsep} 

\definecolor{darkpastelred}{rgb}{0.76, 0.23, 0.13}
\definecolor{ao(english)}{rgb}{0.0, 0.5, 0.0}

\definecolor{darkpastelred}{rgb}{0.76, 0.23, 0.13}
\definecolor{ao(english)}{rgb}{0.0, 0.5, 0.0}

\definecolor{yellow}{RGB}{255,255,153}
\definecolor{grey}{RGB}{224,224,224}

\newboolean{showcomments}
\setboolean{showcomments}{true}
\ifthenelse{\boolean{showcomments}}
 { \newcommand{\mynote}[2]{
      \fbox{\bfseries\sffamily\scriptsize#1}
        {\small$\blacktriangleright$\textsf{\emph{#2}}$\blacktriangleleft$}}}
        { \newcommand{\mynote}[2]{}}

\definecolor{DarkOrange}{rgb}{0.8,0.3,0.0}
\definecolor{DarkCyan}{rgb}{0.0, 0.55, 0.55}
\definecolor{DarkCyel}{rgb}{1.0, 0.49, 0.0}
\definecolor{yellow-green}{rgb}{0.6, 0.8, 0.2}

\newcolumntype{?}{!{\vrule width 1pt}}

\newcommand{\etal}{\emph{et~al.}\xspace}
\newcommand*{\ie}{i.e., }
\newcommand*{\eg}{e.g., }

\setcopyright{none}
\renewcommand\footnotetextcopyrightpermission[1]{} 
\begin{document}
\title{Is ChatGPT the Ultimate Programming Assistant - How far is it?}

\author{Haoye Tian}
\email{haoye.tian@uni.lu}
\affiliation{%
   \institution{University of Luxembourg}
}

\author{Weiqi Lu}
\email{wluak@connect.ust.hk}
\affiliation{%
   \institution{The Hong Kong University of Science and Technology}
}

\author{Tsz-On Li }
\email{toli@connect.ust.hk}
\affiliation{%
   \institution{The Hong Kong University of Science and Technology}
}

\author{Xunzhu Tang}
\email{xunzhu.tang@uni.lu}
\affiliation{%
   \institution{University of Luxembourg}
}

\author{Shing-Chi Cheung}\authornote{Corresponding author.}
\email{scc@cse.ust.hk}
\affiliation{%
   \institution{The Hong Kong University of Science and Technology}
}

\author{Jacques Klein}
\email{jacques.klein@uni.lu}
\affiliation{%
  \institution{University of Luxembourg}
}

\author{Tegawendé F. Bissyandé}
\email{tegawende.bissyande@uni.lu}
\affiliation{%
  \institution{University of Luxembourg}
}

\begin{abstract}
Recent advances in generative artificial intelligence (AI) techniques have significantly impacted software engineering, with AI-driven approaches addressing typical developer challenges, including synthesizing code from descriptions,  repairing faulty programs or providing natural language summaries of existing programs. In particular, large-scale language models (LLMs), which are now increasingly adopted have proven useful in the realm of AI-driven software engineering, as exemplified by OpenAI's Codex. Recently, the ChatGPT LLM has received great attention: it can be used as a bot for discussing source code, prompting it to suggest changes, provide descriptions or even generate code. Typical demonstrations generally focus on existing benchmarks, which may have been used in model training (i.e., data leakage). To assess the feasibility of using an LLM as a useful assistant bot for programmers, we must assess its realistic capabilities on unseen problems as well as its capabilities on various tasks. 
In this paper, we present an empirical study of ChatGPT's potential as a fully automated programming assistant, focusing on the tasks of code generation, program repair, and code summariziation. The study investigates ChatGPT's performance on common programming problems and compares it with state-of-the-art approaches on two benchmarks. Among several findings, our study shows that ChatGPT is effective in dealing with common programming problems. However, our experiments also reveal limitations in terms of its attention span: detailed descriptions will constrain the focus of ChatGPT and prevent it from leveraging its vast knowledge to solve the actual problem. Surprisingly, we have identified the ability of ChatGPT to reason the original intention of the code. We expect future work to build on this insight for dealing with the open question of the oracle problem. 
Our findings contribute interesting insights to the development of LLMs for programming assistance, notably by demonstrating the importance of prompt engineering, and providing a better understanding of ChatGPT's practical applications for software engineering.
\end{abstract}



\maketitle


\section{Introduction}
\label{sec:intro}

Artificial intelligence (AI) has shown great potential in contributing to automate the execution of several challenging software engineering tasks, 
such as code generation~\cite{svyatkovskiy2020intellicode, li2022competition, siddiq2022empirical, bavishi2019autopandas}, program repair~\cite{gupta2017deepfix, ye2022neural, li2022dear, jiang2023knod}, and code summarization/explanation~\cite{hu2018deep, li2020deepcommenter, wang2020reinforcement, stapleton2020human}. Indeed, recent state of the art approaches in code generation leverage AI models to automatically produce full or partial programs based on natural language descriptions or some code inputs. In the program repair realm, neural networks based model are becoming de facto standard for learning to generate fixes, thus reducing manual debugging efforts and enhancing software reliability and security. To address code summarization, researchers have employed generative AI models that output natural language descriptions given an input code, towards contributing to facilitating code comprehension, code maintainability, and code reuse.

The recent emergence of large-scale language models (LLMs) has received much attention in the society in general. In software engineering, it has brought AI-driven automation  to new heights~\cite{devlin2018bert, chen2021evaluating, zhang2020sentiment, tian2020evaluating}. LLMs, which are pre-trained on vast amounts of source code and natural language data, have demonstrated exceptional capabilities in understanding code structures and generating code or texts. Advances led by LLMs have further enhanced the effectiveness of automatic techniques for various software engineering challenges, such as code generation~\cite{nijkamp2022conversational, fried2022incoder, tunstall2022natural, dakhel2022github, chakraborty2022natgen}, program repair~\cite{zhang2022repairing, jain2022jigsaw, jiang2023impact, dakhel2022github}, and code summarization~\cite{ahmed2022few, hu2020deep}. A notable example of LLM is OpenAI's Codex~\cite{chen2021evaluating} (the model behind Github Copilot~\cite{copilot}). Codex has been successfully applied to a range of software engineering tasks, demonstrating its ability to generate accurate and efficient code snippets~\cite{yetistiren2022assessing, vaithilingam2022expectation}, identify and fix bugs or vulnerabilities in software~\cite{pearce2021can, prenner2021automatic}, and provide summarization for code segments~\cite{sarsa2022automatic, macneil2023experiences}. Literature experimental results showcase the enormous potential of LLMs in the software development community although the overall achieved performance remains relatively limited~\cite{chen2022codet, fan2022automated}.

ChatGPT~\cite{ChatGPT}, a recently introduced LLM, has generated significant interest within the software engineering community due to its reported capabilities in the longstanding dream of software engineering: to repair software automatically with minimal human intervention~\cite{sobania2023analysis, xia2023conversational, surameery2023use}. Reported results suggest that ChatGPT will be transformative for the field, and that LLM-driven software engineering has a bright future. We posit that further investigations are still needed to clearly scope the capabilities of LLMs. Indeed, existing program repair studies with ChatGPT tend to assess its performance using old publicly-available benchmark data (anterior to 2022), which may have been leaked into the training corpus of ChatGPT. This experimental bias potentially jeopardizes the generalizability of reported results to new and unseen problems. Furthermore, ChatGPT's code generation and summarization/explanation capabilities have not been thoroughly investigated in the literature. 

To fill these gaps, we propose to carry out an extensive empirical study to evaluate ChatGPT's realistic potential as a fully automated programming assistant. Our study focuses on the key challenging tasks of code generation, program repair, and code summarization, as these tasks are not only repetitive but also fundamental to programmers' daily work~\cite{kazemitabaar2023studying, becker2023programming, xu2022ide, le2012systematic, yang2019survey, rodeghero2014improving, cordy2003comprehending}. Our experimental setup targets common programming problems which are recurrently encountered by programmers~\cite{capilla2019opportunities}. We first evaluate the natural language-to-code (code generation) ability of ChatGPT on two LeetCode programming datasets. Second, we investigate ChatGPT as a program repair tool to repair a large set of diverse incorrect codes in a programming assignments benchmark. Since assignments are usually common coding problems and they are segments of code, the findings made on these assignments are likely to be generalizable~\cite{yang2019refactory, le2015manybugs}. 
Finally, we explore whether ChatGPT can accurately explain the intention of the proposed correct and incorrect code samples in the student assignment benchmark. We will elaborate on how we mitigate the issue of data leakage in the experiments.

\begin{table}[!t]
\centering
\caption{Summary of findings and actionable insights from the experiments assessing the ability of ChatGPT on code generation, program repair and code summarization.}
\label{tab:table_findings}
\small

\begin{tabular}{|p{6.5cm}|p{6.5cm}|} 
\hline
\rowcolor{gray!50}
Findings on Code Generation (Section~\ref{subsec:rq1}) & Actionable insights \\
\hline
\ding{182} ChatGPT is capable of generating correct code across various levels of difficulty and for a wide range of types of problems from the LeetCode benchmark. It outperforms the prior state-of-the-art approaches. &
\ding{172} Since LeetCode repertories common programming problems, our study insight is that ChatGPT can be recommended to developers as an artificial assistant for generating solutions for common and recurrent programming questions.\\
\hline
\ding{183} ChatGPT, like prior LLMs, struggles to generalize its code generation capabilities to new and unseen problems. In the LeetCode benchmark, ChatGPT performs drops significantly for medium and hard problems. &
\ding{173} Developers should not rely on ChatGPT as an assistant bot when dealing with new problems, especially those that are on a higher-level of difficulty.
LLM research should also address the ability to abstract away problem descriptions in order to address generalization (i.e., the cold start problem). \\
\hline
\ding{184} Long description inputs (i.e, lengthy prompts) appear to have a negative impact on the code generation of  ChatGPT and Codex. &
\ding{174} Prompt engineering is a key challenge that developers must address in order to harness the full capability of LLMs. Ensuring the quality (concision and precision) of problem descriptions will facilitate developers engaging with ChatGPT to generate code within a single dialogue session. \\
\hline
\end{tabular}

\vspace{1mm}
\begin{tabular}{|p{6.5cm}|p{6.5cm}|} 
\hline
\rowcolor{gray!50}
Findings on Program Repair (Section~\ref{subsec:rq2}) & Actionable insights \\
\hline
\ding{185} ChatGPT achieves competitive results compared with the state-of-the-art semantics-based Refactory approach, yielding an 84\% success rate (i.e., at least one among TOP-5 outputs is correct) in repairing common programming assignments. However, its average repair ability (i.e., the ratio of correct outputs among the five outputs) remains relatively low at 60\%. &
\ding{175} Enhancing diversity (randomness) in the output of code generation tools augments the likelihood of discovering the correct solution. ChatGPT can be used as a bot assistant to which one must request several code suggestions. A large number of outputs are wrong. 
This suggests that the LLM must be used only by developers who can also assess the output. ChatGPT is a helper and not an autonomous programmer bot, yet. \\
\hline 
\ding{186} Providing a problem description for a buggy code does not guarantee better performance in ChatGPT's repair. If the description lacks precise formal details (e.g., problematic inputs, bug specifications such as test cases or explicit execution requirements), the ChatGPT's ability to fix incorrect code may be reduced since its attention appears to be drawn to non-bug-related details in the description. &
\ding{176} ChatGPT has a limited attention span. Programmers are recommended to provide specific details about the buggy code to fully activate the code-to-code capability of ChatGPT. Prompt engineering for software engineering constructs is again critical.\\
\hline
\end{tabular}

\vspace{1mm}
\begin{tabular}{|p{6.5cm}|p{6.5cm}|} 
\hline
\rowcolor{gray!50}
Findings on Code Summarization (Section~\ref{subsec:rq3}) & Actionable insights \\
\hline
\ding{187} ChatGPT is able to explain the originally intended function (problem description) of code, even when it contains erroneous segments. &
\ding{177} ChatGPT can identify the original intention behind what should be the likely-correct version of the code. This is a key insight that it may be leveraged for tackling the test oracle problem (i.e., predict what should have been the correct behaviour of a given program, opening possibilities for helping to determine the expected input and output and the associated test suites). \\
\hline
\end{tabular}

\end{table}

\paragraph{Contributions}
\begin{itemize}[leftmargin=*]
    \item We conduct an empirical study of ChatGPT on the tasks of code generation, program repair, and code summarization on two benchmarks. 
    \item  We investigate the realistic performance of ChatGPT on common programming problems, comparing it against state-of-the-art approaches.
    \item We make interesting findings of ChatGPT performance and analyze their implications in Table~\ref{tab:table_findings}. The findings contribute actionable insights to the adoption of LLMs for programming assistance and offer a better understanding of ChatGPT's practical applications in the software engineering community. 
\end{itemize}

The remainder of this paper is organised as follows: Section~\ref{sec:bg} presents background details as well as related work. The study design is presented in Section~\ref{sec:sd}. Experimental results are described in Section~\ref{sec:experiments} and threats to validity are enumerated in Section~\ref{sec:threats}. We conclude the paper in Section~\ref{sec:conclusion}.

\section{Background and Related Work}
\label{sec:bg}

\subsection{Large-scale Language Models and ChatGPT}

Large Language Models (LLMs) are a class of deep learning models that have significantly impacted natural language processing (NLP) in recent years with a wide range of applications in various domains~\cite{chen2021evaluating, gu2021domain, alayrac2022flamingo, beltagy2019scibert}. 
Encoder model BERT (Bidirectional Encoder Representations from Transformers)~\cite{devlin2018bert} is a widely-used LLM, built on a transformer-based architecture that uses a Masked Language Modeling (MLM) pre-training task, where it masks some of the input tokens and tries to predict them from the remaining tokens bidirectionally. The MLM is represented as:
\[
L = -\log\left(\frac{e^{s_{\text{target}}}}{\sum_{k}e^{s_k}}\right)
\]
where $s$ is the model's prediction score for the masked word.
This enables BERT to learn context-aware word representations and capture the relationships between words in a sentence. Therefore, Bert, for its ability on code/text understanding, has been applied as an embedding model (encoder) for downstream software engineering tasks~\cite{mashhadi2021applying, ciborowska2022fast, chen2022varclr, tian2020evaluating}.
Decoder model GPT (Generative Pre-trained Transformer)~\cite{ChatGPT, ouyang2022training, chen2021evaluating, radford2019language, radford2018improving}, on the other hand, was trained using the unsupervised learning technique of self-supervised pre-training. It uses a unidirectional, causal attention mechanism that allows each position in the sequence to attend only to preceding positions, preserving the input order. During pre-training, the model aims to maximize the likelihood of predicting a word based on its contextual information:
\[
L = \sum_{t} \log P(w_t | w_{1:t-1}; \theta)
\]
where \( L \) is the likelihood, \( w_t \) is the word at position \( t \), \( w_{1:t-1} \) is the sequence up to \( t-1 \), and \( \theta \) are the model parameters.
GPT-1~\cite{radford2018improving} attempts to predict the next word in a sequence given the previous words. This task is performed unidirectionally from left to right, allowing GPT-1 to learn the dependency relations between words and generate relevant and context-aware content. 
GPT-2~\cite{radford2019language} is a larger model than GPT-1, with 1.5 billion parameters 10 times compared to GPT-1's 117 million parameters. GPT-2 was trained with larger architecture on a larger and more diverse dataset than GPT-1, which includes web pages, books and scientific articles. GPT-2 demonstrated that training on a larger dataset and utilizing a greater number of parameters can enhance the capabilities of a language model even in zero-shot settings.
Compared to GPT-1 and GPT-2, GPT-3~\cite{brown2020language} represents a significant advancement in LLM development. It has a much larger 175 billion parameters. Codex~\cite{chen2021evaluating} is initially created using GPT-3 weights derived from training on a natural language corpus, and then it is fine-tuned on an extensive dataset of code files. Afterwards, the InstructGPT~\cite{ouyang2022training} was secondly created to be fine-tuned on GPT-3 by using reinforcement learning from human feedback (RLHF).

ChatGPT~\cite{ChatGPT} is the most popular ChatBot (GPT-3.5 and GPT-4) developed by OpenAI. It is a sibling model of InstructGPT while they are trained with different training datasets because of the different purposes. The training dataset of InstructGPT mainly consists of instructional text such as explanations, and examples. Specifically, ChatGPT focuses on conversation, thus ChatGPT is trained with massive texts and codes, including a wide range of user queries, and responses which are coherently relevant to these queries. These training data make ChatGPT applicable to chatbots or virtual assistants. Furthermore, it also employs advanced supervised instruction fine-tuning techniques and RLHF to adapt more effectively to specific tasks or domains.
A recent study~\cite{sobania2023analysis} shows that ChatGPT obtains significant improvement over other LLMs. Although OpenAI has not released official information regarding the technical details of ChatGPT, its enhanced performance is likely attributable to its training techniques and large corpus. Since ChatGPT is the most advanced and possibly the most powerful LLM, we conducted an extensive experiment to investigate ChatGPT's capability in performing various software engineering tasks (\S\ref{sec:crt}).


\subsection{Code-related Tasks}\label{sec:crt}

\subsubsection{Code generation.}
Code generation is the process of automatically creating entire program code or completing code snippets from a higher-level representation, such as a natural language description, a model or specification, to improve programming efficiency and reduce human error~\cite{budinsky1996automatic, svyatkovskiy2020intellicode, parvez2021retrieval}. Recent advances in machine learning and natural language processing have led to the development of new techniques for code generation. 
Specifically, Liu~\etal~\cite{liu2020self} proposed a self-attention neural architecture for code completion by utilizing hierarchical structural information, capturing long-term dependency, and employing a multi-task learning framework for knowledge sharing on real-world datasets. 
Wei~\etal~\cite{wei2019code} presented a dual training framework that simultaneously trains code summarization and code generation tasks by exploiting their duality, resulting in improvements for both tasks on GitHub datasets. With the development of pre-trained LLMs, researchers attempt to use them for code generation.

Ahmad~\etal~\cite{ahmad2021unified} proposed PLBART, a sequence-to-sequence model, pre-training on a large dataset and effectively learning syntax, style, and logical flow. InCoder~\cite{fried2022incoder} is a unified generative model capable of program synthesis and editing through infilling, which leverages bidirectional context to improve performance.  Nijkamp~\etal~\cite{nijkamp2022conversational} introduced CodeGen, a family of large language models up to 16.1B parameters trained on natural language and programming data, demonstrating competitive performance on zero-shot Python code generation. 
More recently, Yetistiren~\etal~\cite{yetistiren2022assessing} evaluated the quality of code generated by GitHub Copilot, producing a 91.5\% success rate in generating valid code. However, the results suggest more comprehensive assessment is needed.
In this paper, we explore LLM in generating code functions through the programming problem descriptions in natural language. 

\subsubsection{Program repair.}
Program repair is designed to automatically identify and repair bugs or vulnerabilities in software code~\cite{le2021automatic, motwani2020quality, shariffdeen2021concolic, zhang2022heterogen}. Specifically, Gulwani~\etal~\cite{gulwani2018automated} presented an automated program repair algorithm for introductory programming assignments by using correct student solutions to repair incorrect attempts. Hu~\etal~\cite{yang2019refactory} proposed a search-based fully automated approach for generating real-time student program repairs by refactoring correct solutions and matching incorrect programs to refactored ones. The method requires fewer correct solutions and achieves better results with just a single correct reference solution. On the other hand, AI-based systems have recently achieved state-of-the-art performance in this task. 
Chakraborty~\etal~\cite{chakraborty2020codit} proposed a novel tree-based neural network system, CODIT, to learn and suggest code change patterns from large-scale open-source data. The evaluation demonstrates its effectiveness in suggesting patches and fixing bugs on the Defects4J dataset. 
Ye~\etal~\cite{ye2022neural} proposed a program repair model called RewardRepair which uses program-specific information during neural network weight optimization to produce patches that compile and are not prone to be overfitting. 

With respect to the LLM techniques, Prenner~\etal~\cite{prenner2022can} investigated pre-trained large language model Codex's ability to localize and fix bugs in code, finding that it is surprisingly effective and competitive with state-of-the-art automated program repair techniques, especially in repairing Python bugs. Sobania~\etal~\cite{sobania2023analysis} evaluated ChatGPT's repair performance on the QuixBugs benchmark set. By providing hints through its conversational system, ChatGPT is able to repair 31 out of 40 bugs and outperform state-of-the-art approaches. Meanwhile, Xia~\etal~\cite{xia2023conversational} demonstrated the improved performance of conversational APR over prior LLM-based APR methods. 
In our study, we aim to investigate the automatic repair of diverse incorrect implementations submitted for common programming assignments.

\subsubsection{Code summarization.}
Code summarization refers to the automatic process of providing clear and concise summarizations or explanations for a piece of code or functionality~\cite{hu2018deep, parvez2021retrieval, ahmad2020transformer, shi2022evaluation}. Allamanis~\etal~\cite{allamanis2016convolutional} leveraged a simple Transformer model with self-attention mechanisms for source code summarization that significantly outperforms state-of-the-art techniques. Some researchers proposed to summarize code by leveraging the abstract syntax trees (AST) of code. For instance, Hu~\etal~\cite{hu2018deep} proposed a novel approach that uses a structure-based traversal (SBT) method to flatten the AST into a sequence to automatically generate code comments for Java methods. Shido~\etal~\cite{shido2019automatic} proposed Multi-way Tree-LSTM, an extension of Tree-LSTM, to handle source code summarization by addressing the challenges of applying machine translation models to source code's ASTs with nodes having arbitrary numbers of children and their order. 
Wan~\etal~\cite{wan2018improving} also incorporated an AST structure and sequential content into a deep reinforcement learning framework, using an actor-critic network to optimize word prediction and provide global guidance for exploration. LeClair~\etal~\cite{leclair2020improved} presented a graph-based neural architecture that combines both the source code sequence and structural information from ASTs to improve automatic source code summarization. In our paper, we focus on the summarization and explanation of the intention of the source code.

\section{Study Design}
\label{sec:sd}
In this section, we first present three research questions that we aim to investigate the ChatGPT's ability in this study. Then, we introduce two benchmarks and evaluation metrics, describing how we elaborate them for the research questions. Finally, we present the application of ChatGPT.

\subsection{Research Questions}
\begin{description}
	\item {\em {\bf RQ-1:} To what extent can ChatGPT generate correct and efficient code for common programming problems?} We propose to assess ChatGPT's text-to-code generation ability by measuring the correctness and time complexity of its code generated for common programming problems on two datasets that are built from LeetCode 2016-2020 and LeetCode 2022 respectively.
	\item {\em {\bf RQ-2:} Can ChatGPT repair diverse buggy code implemented for common programming problems effectively?} To answer this research question, we leverage a benchmark of algorithmic programming assignments that contains 1783 diverse incorrect code submissions. We investigate whether ChatGPT can fix the incorrect code with or without the problem descriptions provided.
   	\item {\em {\bf RQ-3:} Can ChatGPT identify the intention of code?} We propose to investigate the ability of ChatGPT to explain code. Specifically, we evaluate whether ChatGPT can infer the original intention of code, even in cases where the code contains errors.
\end{description}

\subsection{Benchmark selection criteria}
Python is a widely used high-level programming language for coding applications. 
In addition, Python is the programming language that descendant models of OpenAI GPT-3 (e.g. ChatGPT and Codex) are most capable in~\cite{OpenAIModels, OpenAIModels2}. 
To investigate the ability of ChatGPT in code-related tasks, we thus adopt Python as the programming language for the benchmarks in our study. 
Furthermore, our study scope requires benchmarks that have a diversity of programming problems as well as implementations.

\subsection{Datasets}
\label{sec:dataset}

We base our study on two benchmarks. 
The first benchmark is built on LeetCode~\cite{LeetCode}, an online platform offering a diverse range of programming problems, such as algorithms, data structures, etc. LeetCode categorizes programming problems based on the types of problems (e.g. array, sorting, etc.) and difficulty level of problems (easy, medium, and hard). LeetCode also provides adequate test cases (avg. 135 per problem) to verify the correctness of submitted code and time-complexity-based rank percentile to evaluate the efficiency of the code. These informative features motivate us to select it as our first benchmark.
In order to ensure the generalizability of our findings, we focus on the four most common (number) types\footnote{We do not select Math (272) and Dynamic programming (267) as they are exclusively appearing in either the easy problem or hard problem respectively.} of programming problems in the entire LeetCode repository: Array (862), String (398), Hash table (283), Sorting (203). We randomly sample 10 problems from each of the 4 problem types for the 3 levels of easy, medium, and hard. Finally, we obtain 120 (10*4*3) sampled programming problems from LeetCode in 2016-2020 and LeetCode in 2022 respectively. We describe how we leverage them at the end of this section.

The second benchmark is Refactory~\cite{yang2019refactory}, a large Python repairing bug benchmark released in 2019. It contains 2442 correct and 1783 buggy programs submitted by undergraduate students at the National University of Singapore for five algorithmic common programming assignments. The various and diverse correct and incorrect Python code contribute to our study. Besides, Refactory also provides instructor-designed test cases (avg. 9 per problem), correct reference solutions and descriptions\footnote{we supplement with detailed descriptions provided in ~\cite{dakhel2022github}} for each assignment. The test suite is designed to provide students with evaluation results as feedback. We take this benchmark into our account as it contains plenty of implementations and well-designed test cases for common programming problems (\eg searching/sorting/deduplication algorithms). Table~\ref{tab:dataset} shows a summary of the benchmark.

\begin{table}
	\centering
	\caption{A summary of the Refactory Benchmark.}
	\label{tab:dataset}
	\resizebox{1\linewidth}{!}
	{
		\begin{tabular}{c|c|c|c}
			\toprule
             {\bf Problem} & {\bf Description} & {\bf \#Correct} & {\bf \#Incorrect} \\
		\hline

            {\makecell[c]{Sequential \\ Search}} & \makecell[l]{This function takes in a value “x” and a sorted sequence  \\ “seq”, and returns the position that “x” should  go to suc-  \\ h that the sequence remains sorted. Otherwise,  return t- \\ he length of the sequence.} & \makecell[c]{768} & \makecell[c]{575}   \\
            \hline

            {\makecell[c]{Unique Dates \\ Months}} & \makecell[l]{Given a month and a list of possible birthdays, returns \\ True if there is only one possible birthday with that m- \\  onth and unique day, and False otherwise.  Implement 3\\  
            different functions: unique\_day, unique\_month and co-\\ ntains\_unique day.} & \makecell[c]{291} & \makecell[c]{435}   \\
            \hline

            {\makecell[c]{Duplicate \\ Elimination}} & \makecell[l]{This function takes in a list and returns a new list with \\ all repeated occurrences  of any element removed.} 
             & \makecell[c]{546} & \makecell[c]{308}   \\
            \hline

            {\makecell[c]{Sorting \\ Tuples}} & \makecell[l]{Given a list of people, this function sorts the people an- \\ d returns a list in an order such that the older people a-\\ re at the front of the list.} & \makecell[c]{419} & \makecell[c]{357}   \\
            \hline

            {\makecell[c]{Top\_k \\ Elements}} & \makecell[l]{This function top\_k accepts a list of integers as the inp- \\ ut and returns the greatest k number of values as a list, \\  with its elements sorted in descending order.} & \makecell[c]{418} & \makecell[c]{108}   \\
            \hline
            Total & & 2442 & 1783 \\
		   \bottomrule
		\end{tabular}
	}
\end{table}

We note that ChatGPT has been trained on a diverse and extensive dataset of text, including sources such as web pages, books, articles, and online forums. {\bf This actually presents a potential data leakage problem for the relevant research. Because when selecting experimental benchmarks on which ChatGPT will be evaluated, we are uncertain whether ChatGPT has already learned the solution references provided in these benchmarks in advance.} In this case, the achieved performance and insights about ChatGPT may not be able to be generalized to new or unseen problems. 
To evaluate ChatGPT's true capability and mitigate the possible cold start problem\footnote{The model cannot draw any inferences for items about which it has not yet gathered sufficient information.} of ChatGPT, we elaborate on how we utilize the two benchmarks mentioned above for each research question. We do not aim to construct entirely new programming problems for ChatGPT; expecting so would be unrealistic. Similar information exists widely online, and this existing knowledge is essential for training language models like ChatGPT to effectively predict and address future issues.

\begin{itemize}[leftmargin=*]
    \item {\bf RQ-1: Code generation.} Refactory only contains five programming problems and all of them can be accomplished by ChatGPT and Codex (). Therefore, we mainly use LeetCode programming problems to evaluate the code generation ability of ChatGPT. Note that the knowledge cutoff of ChatGPT is {\em September 2021}. We first evaluate ChatGPT on 120 (10*4*3) programming problems from LeetCode in 2016-2020 that ChatGPT may have learned knowledge from. Then, to investigate whether ChatGPT can generalize its ability to new problems, we assess ChatGPT on another 120 (10*4*3) programming problems from LeetCode in 2022. We skip LeetCode in 2021 as it's mixed of potentially trained problems and new problems for ChatGPT and we wish to evaluate ChatGPT separately on two kinds of datasets.
    \item {\bf RQ-2: Program repair.} To evaluate ChatGPT's ability to program repair, we utilize the 1783 buggy programs of five programming problems in Refactory benchmark. The selection is based on: (1) The repair solutions of buggy programs are not publicly available, and therefore, ChatGPT is more likely not to have been trained on them in advance. (2) The buggy assignment code are diverse and implemented for common coding problems (searching/sorting algorithms), and they are segments of code, the findings made on these assignments are likely to be generalizable~\cite{yang2019refactory, le2015manybugs}. Although ChatGPT can access the five teacher-written correct reference solutions, it is not an issue for such an assignment repair tool (ChatGPT's role in this research question) that is allowed to learn from the correct reference solutions.
    \item {\bf RQ-3: Code summarization.} We assign ChatGPT the task of generating code summarization for the intention of various correct and incorrect codes. Therefore, we seek 2442 correct and 1783 buggy codes in Refactory. There are no natural language explanations provided for them on the Internet. We can be relatively safe to request ChatGPT to produce the natural language intention of code.
\end{itemize}

\subsection{Application of ChatGPT}
\label{chatgpt_application}
ChatGPT was a newly-released LLM-based conversational product by OpenAI in November 2022. There has not been a consensus on the standard protocol for the usage of ChatGPT. In this section, we describe how we apply ChatGPT in our experiments.

\begin{enumerate}[leftmargin=*]
\item {\bf ChatGPT API.} OpenAI has introduced the ChatGPT API for several of its developed models, including GPT-3 and GPT-4. Notably, the GPT-4 model—the latest in the series—comes with aggressive request rate limits~\footnote{https://platform.openai.com/docs/guides/rate-limits/overview}, posing challenges for large-scale experimentation. Concurrently, a surge in recent reports and studies suggests that GPT-4 is exhibiting signs of {\bf instability} and inaccuracy~\cite{noller2023gpt4analysis, piunikaweb2023chatgpt, businessinsider2023gpt4, chen2023chatgpt}, which may stem from its radical redesign. In contrast, the GPT-3.5 family produce reliable results and has been widely used. Given this context, we have chosen the GPT-3.5 model for our investigation. Specifically, our research employs the {\em gpt-3.5-turbo-0613} model with its default parameter configurations.


\item {\bf Randomness.} Given the inherent randomness of GPT models, it is important to note that ChatGPT has the possibility to generate different responses for the same prompt input at different requests.
In particular, when it comes to code generation and program repair, ChatGPT may attempt different data structures or algorithms to achieve these tasks. To deal with randomness and ensure reliable analysis and conclusion, there is a need to request multiple times to ChatGPT for the same prompt. 
We follow the prior works and run five independent conversational requests with ChatGPT for each prompt. We display the performance in two measurements: 
(1) {\bf TOP-5}, the value is 1 if at least one of the five attempts of an approach solves the specific programming problem, otherwise 0. This metric indicates the robustness of an approach. Specifically, it assesses the presence or absence of a viable solution within the top five attempts, which, according to a survey~\cite{noller2022trust}, is the maximum number of generated code that most developers are willing to review.
(2) {\bf AVG-5}, the average success rate of five attempts of an approach to the programming problem. A score of 1 for AVG-5 means that all five attempts are successful. Given the inconsistency/randomness of the TOP-1 performance of LLMs, we therefore propose the AVG-5 to represent the average TOP-1 of an approach. A high AVG-5 score indicates that the approach consistently yields correct solutions, making it a reliable tool for programmers in need of precise outcomes.


\item {\bf Prompt.} The prompt is a critical component in utilizing ChatGPT and other LLMs, as it provides the context and direction for the generation of texts or codes.
We introduce the baseline prompt we apply for the research experiments.
First, Figure~\ref{fig:code_generation} shows an example of a prompt that we request ChatGPT for the code generation task in RQ-1. The prompt is mainly composed of the programming task description given by LeetCode. The description comprises a programming problem introduction (lines 1: natural language description, lines 3-9: examples, lines 11-12: constraints) and a defined function to be completed (lines 14-17). To form the final prompt for Python code generation, we append a fixed request statement, "Implement the above task in Python." at the end.
For the task of program repair (RQ-2), OpenAI set up a prompt template that we follow up on for our experiments, as shown in Figure~\ref{fig:program_repair}. For the task of code summarization (RQ-3), we recall that our expectation is to obtain the explanation of ChatGPT about the semantic intention of code without describing what each line of code is doing. Therefore, we construct our prompt as to be the fixed request statement: ``Can you explain the intention of the below function(s) within one sentence? Do not include any explanations of code details in your answer.'' followed by the associated code.
\end{enumerate}

\begin{figure}[ht]
    \centering
    \normalsize
\begin{lstlisting}[language=Text,linewidth={\linewidth},numbers=none,frame=tb,basicstyle=\ttfamily,basicstyle=\normalsize]
(@\bf 1@)   Convert a non-negative integer `num` to its English words representation.
(@\bf 2@)
(@\bf 3@)   **Example 1:**
(@\bf 4@)      Input: num = 123    Output: "One Hundred Twenty Three"
(@\bf 5@)   **Example 2:**
(@\bf 6@)      Input: num = 12345    Output: "Twelve Thousand Three Hundred Forty Five"
(@\bf 7@)   **Example 3:**
(@\bf 8@)      Input: num = 1234567    Output: "One Million Two Hundred Thirty Four Thousand 
(@\bf 9@)      Five Hundred Sixty Seven" 
(@\bf 10@)
(@\bf 11@)   **Constraints:**
(@\bf 12@)     * `0 <= num <= 2^31 - 1`
(@\bf 13@)
(@\bf 14@)   ```
(@\bf 15@)   class Solution:
(@\bf 16@)        def numberToWords(self, num: int) -> str:
(@\bf 17@)   ```
(@\bf 18@)   
(@{\bf 19}@)   (@{\bf{Implement the above task in Python.}}@)
\end{lstlisting}
    \vspace{-3mm}
    \caption{A prompt example of ChatGPT for code generation.}
    \label{fig:code_generation}
\end{figure}

\begin{figure}[ht]
    \centering
    \normalsize
\begin{lstlisting}[language=Text,linewidth={\linewidth},numbers=none,frame=tb,basicstyle=\ttfamily,basicstyle=\normalsize,]
(@\bf 1@)   (@\bf{\#\#\#\#\# Fix bugs in the below function(s)}@)
(@\bf 2@)   (Description)
(@\bf 3@)   (@{\bf{\#\#\# Buggy Python}}@)
(@\bf 4@)   def remove_extras(lst):
(@\bf 5@)       t=[]
(@\bf 6@)       for i in lst:
(@\bf 7@)           if i not in t:
(@\bf 8@)               t.append(i)
(@\bf 9@)   
(@\bf 10@)   (@\bf{\#\#\#} Fixed Python@)

\end{lstlisting}
    \vspace{-3mm}
    \caption{A prompt example of ChatGPT for program repair.}
    \label{fig:program_repair}
\end{figure}

\subsection{Evaluation Metrics}
\label{metrics}

\begin{itemize}[leftmargin=*]
    \item {\bf Code correctness.} To assess the correctness of generated code in RQ-1 and RQ-2, we rely on test cases that the benchmark (LeetCode) provides. Any code that passes all test cases is considered correct, while any code that fails any test case is considered incorrect.
    We then use TOP-5 and AVG-5 metrics as overall performance measures, as detailed in the last section (item (2) Randomness). 
    \item {\bf Code time complexity.} For the RQ-1 task of code generation, we also pay attention to the efficiency of the code. We rely on the official time-complexity-based rank percentile of code submitted to LeetCode as a key metric to evaluate the performance of different code generation techniques and benchmark them against human-written code. 
    More specifically, we compute the average time-complexity-based rank percentile. If the average rank is, e.g.,  10\%, this means that the generated code is among the top 10\% of the most efficient solutions.
    \item {\bf Repair rate.} To evaluate the performance of automated program repair tools in RQ-2, we leverage the standard metric of repair rate, i.e. the ratio of the patched code generated by approaches that pass all instructor-designed test cases.  
    We use TOP-5 and AVG-5 metrics as performance measures, as detailed in the last section (item (2) Randomness). 
\end{itemize}

\section{Experiments \& Results}
\label{sec:experiments}

In this section, we present the experiments performed to answer our research questions. In Sections~\ref{subsec:rq1} and \ref{subsec:rq2}, we investigate the performance of ChatGPT in generating code and repairing buggy programs respectively. In Section~\ref{subsec:rq3}, we examine whether ChatGPT can accurately explain the intention of code.

\subsection{ChatGPT for Code Generation}
\label{subsec:rq1}

\noindent
{\bf [Experiment Goal]:} We aim to evaluate ChatGPT's text-to-code generation ability for common programming problems by assessing its performance on two Leetcode datasets.

\noindent
{\bf [Experiment Design]:} 
As outlined in Section~\ref{sec:dataset}, we utilize the LeetCode datasets from 2016-2020 and 2022 to assess ChatGPT's performance. The programming problem introduction consists of three parts: problem description, examples, and constraints. The problem description is a necessary block. We include examples because there may exist some edge cases without clear specifications explained in the description\footnote{https://leetcode.com/problems/determine-if-two-events-have-conflict/}. The constraints specify the value range and format of the input data, which may not be missing in the description\footnote{https://leetcode.com/problems/minimum-total-cost-to-make-arrays-unequal/}. We include these components in our prompt to ensure the completeness of requirement specifications. 
Regarding the metrics employed, we do not measure the ability of code completion since we have observed that ChatGPT and other approaches can generate code responses for all given problems. Our experiment mainly assesses ChatGPT's performance with respect to the correctness of the generated solutions and the rank percentile of time complexity for the correct solutions, as determined by test suites and execution time calculations within the LeetCode grading system. We engage ChatGPT in code generation tasks with the prompt illustrated in Figure~\ref{fig:code_generation}.
To compare ChatGPT's performance with two state-of-the-art code generation tools, we select: (1) Codex~\cite{chen2021evaluating}, an advanced GPT based LLM model from the OpenAI family that has been widely investigated and applied in software engineering tasks before~\cite{yetistiren2022assessing, vaithilingam2022expectation, pearce2021can, prenner2021automatic, sarsa2022automatic, macneil2023experiences};. When evaluating Codex, we employ the same prompt as ChatGPT and keep the default parameters, with the exception of setting {\em max\_tokens} to 1024. This adjustment of {\em max\_tokens} ensures that we receive the complete code output from Codex. (2) CodeGen~\cite{nijkamp2022conversational}, an autoregressive transformer model outside the OpenAI family. We employ the same prompt and parameters as ChatGPT and Codex for a fair comparison. CodeGen serves as a baseline model outside the OpenAI family for evaluating the performance of ChatGPT in code generation. 
Finally, to explore the usage of LLMs for programmers, we further analyze the impact of the length of prompts (problem description) on the code generation performance of the approaches. We compute the boxplot distribution of the length of prompts for correct and incorrect predictions made by ChatGPT and Codex respectively on LeetCode 2022. 


\noindent
{\bf [Experimental Results]:} 
Table~\ref{tab:code_generation_leetcode_2016-20} demonstrates that ChatGPT outperforms Codex and CodeGen in generating correct code for problems across all difficulty levels and problem types within the LeetCode 2016-2022 dataset. Remarkably, ChatGPT can generate the correct code for {\bf all} selected easy problems and 65\% (26/40) of the hard problems on the metric of TOP-5. However, for AVG-5, we note that ChatGPT can generate correct code for 43\% of hard problems, which indicates requesting ChatGPT multiple times may improve the probability to discover correct solutions.
On the other hand, regarding the time complexity of generated code, the overall rank percentile of ChatGPT-generated correct code surpasses that of Codex-generated correct code while both of them are basically lower 50\% (i.e., always in the top 50\%). These rank scores imply that the efficiency of code produced by ChatGPT and Codex is potentially superior to most human-written code (assuming that nearly all LeetCode submissions are from humans) in terms of time complexity.


Table~\ref{tab:code_generation_leetcode_2022} presents the performance of the prior three approaches on LeetCode 2022 dataset. It shows that both metrics of the correctness and time complexity drop across all difficulty levels and types, with greater reduction as the difficulty increases. These results suggest ChatGPT and other LLMs struggle to generalize their abilities to new and unseen problems, i.e., the cold start problem: The model cannot draw any inferences for problems about which it has not yet gathered sufficient information. Fortunately, when mitigating the impact of data leakage, the results demonstrate that ChatGPT indeed improves the performance and generalization over the prior state of the arts. It can solve most of the easy problems (33/40) and merely 2 hard problems. On the other hand, the efficiency rank of code generated by ChatGPT are still in the top 50\% for easy and medium problems while not for hard problems.


\begin{table}[ht]
	\centering
	\caption{Performance of ChatGPT (GPT), Codex (Dex), and CodeGen (Gen) on code generation for the problems in LeetCode 2016-2020.}
	\label{tab:code_generation_leetcode_2016-20}
		\resizebox{.9\linewidth}{!}
	{
	\begin{threeparttable}
		\begin{tabular}{l|l|rrr|rrr}
			\toprule
			\multirow{2}{*}{\bf Level} & \multirow{2}{*}{\bf Type} & \multicolumn{3}{c|}{\bf Correctness} & \multicolumn{3}{c}{\bf Average Rank }\\\cline{3-8}
			& & {\bf GPT} & {\bf Dex} & {\bf Gen} & {\bf GPT} & {\bf Dex} & {\bf Gen} \\
			\hline
				\multirow{4}{*}{Easy}
			& Array & 10 (0.96) & 10 (0.28) & 2 (0.04) & 8\% & 43\% & 37\% \\
			& String & 10 (0.94) & 7 (0.24) & 2 (0.04) & 10\% & 35\% & 36\% \\
			& Hash Table & 10 (1.00) & 7 (0.36) & 3 (0.14) & 12\% & 21\% & 23\% \\
			& Sorting & 10 (0.92) & 8 (0.42) & 2 (0.06) & 7\% & 20\% & 10\% \\
			\hline
			Total &  & 40 (0.95) & 32 (0.33) & 9 (0.07)  & 9\% & 31\% & 26\% \\
			\hline
				\multirow{4}{*}{Medium}
			& Array & 10 (0.76) & 5 (0.18) & 0 (0.00) & 8\% & 34\% &  -  \\
			& String & 9 (0.72) & 5 (0.16) & 1 (0.02) & 11\% & 49\% & 82\% \\
			& Hash Table & 10 (0.76) & 5 (0.20) & 1 (0.02) & 13\% & 25\% & 89\% \\
			& Sorting & 10 (0.94) & 5 (0.14) & 1 (0.02) & 11\% & 36\% & 27\% \\
			\hline
			Total &  & 39 (0.80) & 20 (0.17) & 3 (0.02)  & 11\% & 36\% & 66\% \\
			\hline
				\multirow{4}{*}{Hard}
			& Array & 6 (0.46) & 3 (0.08) & 0 (0.00) & 26\% & 29\% &  -  \\
			& String & 8 (0.48) & 2 (0.04) & 0 (0.00) & 29\% & 56\% &  -  \\
			& Hash Table & 7 (0.42) & 3 (0.10) & 0 (0.00) & 37\% & 42\% &  -  \\
			& Sorting & 5 (0.36) & 1 (0.02) & 0 (0.00) & 12\% & 15\% &  -  \\
			\hline
			Total &  & 26 (0.43) & 9 (0.06) & 0 (0.00)  & 27\% & 38\% &  -  \\
			\bottomrule
		\end{tabular}
		{$^\ast$ The integer values in column `Correctness' indicate the number of problems for which the approaches can generate correct solutions within five times attempts (TOP-5) given ten programming problems at each type. The decimal values in parentheses indicate the overall average success rate at five times attempts (AVG-5). See randomness in Section~\ref{chatgpt_application}.} \\
			{$^\ast$ The percentage values in column `Average Rank' represent the average rank percentile of generated correct solutions. The lower, the better.}

		\end{threeparttable}
	}
\end{table}

\begin{table}[ht]
	\centering
	\caption{Performance of ChatGPT (GPT), Codex (Dex), and CodeGen (Gen) on code generation for the problems in LeetCode 2022.}
	\label{tab:code_generation_leetcode_2022}
		\resizebox{.9\linewidth}{!}
	{
	\begin{threeparttable}
		\begin{tabular}{l|l|rrr|rrr}
			\toprule
			\multirow{2}{*}{\bf Level} & \multirow{2}{*}{\bf Type} & \multicolumn{3}{c|}{\bf Correctness} & \multicolumn{3}{c}{\bf Average Rank }\\\cline{3-8}
			& & {\bf GPT} & {\bf Dex} & {\bf Gen} & {\bf GPT} & {\bf Dex} & {\bf Gen} \\
			\hline
				\multirow{4}{*}{Easy}
			& Array & 8 (0.74) & 4 (0.12) & 1 (0.02) & 7\% & 18\% & 18\% \\
			& String & 10 (0.50) & 2 (0.10) & 0 (0.00) & 19\% & 27\% &  -  \\
			& Hash Table & 7 (0.58) & 3 (0.08) & 1 (0.02) & 7\% & 42\% & 45\% \\
			& Sorting & 8 (0.52) & 2 (0.10) & 1 (0.04) & 16\% & 21\% & 19\% \\
			\hline
			Total &  & 33 (0.58) & 11 (0.10) & 3 (0.02)  & 13\% & 27\% & 27\% \\
			\hline
				\multirow{4}{*}{Medium}
			& Array & 1 (0.08) & 1 (0.02) & 0 (0.00) & 26\% & 17\% &  -  \\
			& String & 1 (0.10) & 0 (0.00) & 0 (0.00) & 8\% &  -  &  -  \\
			& Hash Table & 5 (0.42) & 2 (0.08) & 0 (0.00) & 40\% & 21\% &  -  \\
			& Sorting & 3 (0.12) & 0 (0.00) & 0 (0.00) & 19\% &  -  &  -  \\
			\hline
			Total &  & 10 (0.18) & 3 (0.03) & 0 (0.00)  & 29\% & 19\% &  -  \\
			\hline
				\multirow{4}{*}{Hard}
			& Array & 0 (0.00) & 0 (0.00) & 0 (0.00) &  -  &  -  &  -  \\
			& String & 2 (0.04) & 0 (0.00) & 0 (0.00) & 71\% &  -  &  -  \\
			& Hash Table & 0 (0.00) & 0 (0.00) & 0 (0.00) &  -  &  -  &  -  \\
			& Sorting & 0 (0.00) & 0 (0.00) & 0 (0.00) &  -  &  -  &  -  \\
			\hline
			Total &  & 2 (0.01) & 0 (0.00) & 0 (0.00)  & 71\% &  -  &  -  \\
			\bottomrule
		\end{tabular}
		{$^\ast$ The integer values in column `Correctness' indicate the number of problems for which the approaches can generate correct solutions within five times attempts (TOP-5) given ten programming problems at each type. The decimal values in parentheses indicate the overall average success rate at five times attempts (AVG-5). See randomness in Section~\ref{chatgpt_application}.} \\
			{$^\ast$ The percentage values in column `Average Rank' represent the average rank percentile of generated correct solutions. The lower, the better.}

		\end{threeparttable}
	}
\end{table}

\begin{figure}[ht]
    \centering
    \includegraphics[width=.6\columnwidth]
    {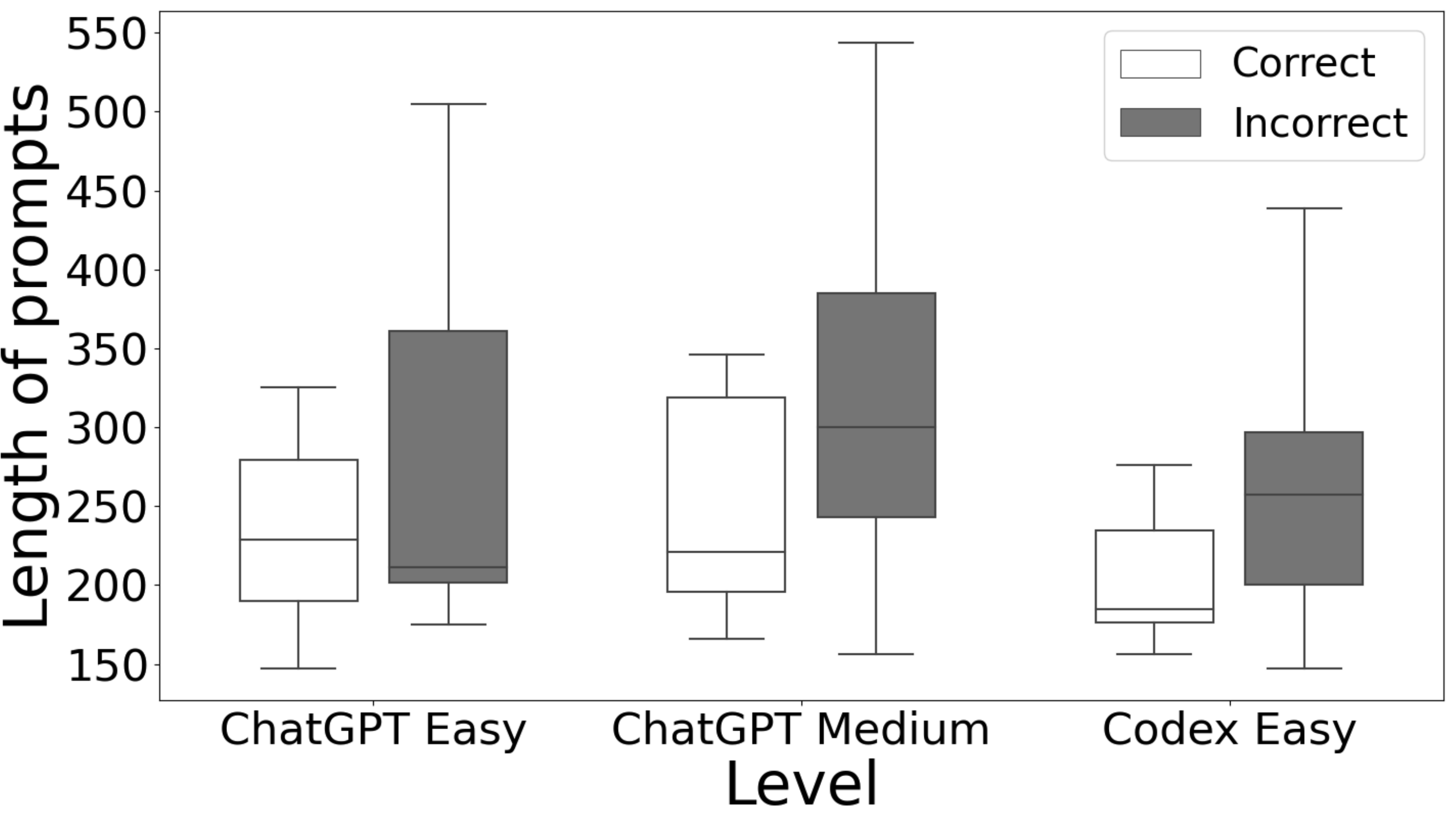}
    \vspace{-1mm}
    \caption{Impact of length of prompts to predictions of ChatGPT and Codex}
    \label{fig:desc_len_mixed}
\end{figure}

Figure~\ref{fig:desc_len_mixed} presents the 
distributions of prompt lengths for the correct and incorrect prediction. We exclude the hard level of ChatGPT and the medium and hard levels of Codex and CodeGen from the analysis since the number of problems solved by them is not sufficient to achieve statistical significance. We 
observe that the length of prompts are 
smaller in the correct predictions than that in the incorrect predictions: This implies that the models make more correct predictions when the prompts have a relatively shorter length. Such a finding has already been demonstrated in the literature~\cite{li-liang-2021-prefix}. Our results suggest that programmers should provide a clear and concise prompt when using ChatGPT and Codex for code generation.

\subsection{ChatGPT for Program Repair}
\label{subsec:rq2}

\noindent
{\bf [Experiment Goal]:} We investigate ChatGPT's ability to repair diverse incorrect code implemented for common programming problems.

\noindent
{\bf [Experiment Design]:} 
As discussed in Section~\ref{sec:dataset}, we attempt to repair a large-scale dataset of 1783 incorrect codes of five common problems in Refactory benchmark with the goal of alleviating data leakage. We design two experiments for the repair of ChatGPT. First, we provide plain buggy code to request ChatGPT to repair by following the prompt shown in Figure~\ref{fig:program_repair}. 
We note that Sobania~\etal~\cite{sobania2023analysis} investigated that ChatGPT can repair more bugs if more information about bugs are provided at multi-round dialogues. 
However, the bug related information is not often available in practice and the multi-round dialogue tends to be time costly. Therefore, to consider the scenarios practical, we in this experiment aim to explore the repair ability of ChatGPT by providing the often available problem (function) description (\ref{tab:dataset}) at a one-time dialogue request. We construct the second prompt by inserting the description of each problem into line 2 of the prior prompt (Figure~\ref{fig:program_repair}). 
For comparison with state-of-the-art, we select two approaches: (1) Codex~\cite{chen2021evaluating}, an advanced GPT based LLM model from the OpenAI family that has been widely investigated and applied in software engineering tasks before~\cite{yetistiren2022assessing, vaithilingam2022expectation, pearce2021can, prenner2021automatic, sarsa2022automatic, macneil2023experiences}; (2) Refactory~\cite{yang2019refactory} (same name as the Refactory benchmark), a semantic-based APR tool for repairing
student programs in real-time by refactoring correct reference solutions. This tool retrieves the top-5 closest programs to construct patches. If any of them pass the test suites, they succeed in repairing the buggy code. We thus compare them with the metric of TOP-5.
We use repair rate to measure approaches' performance while not considering time cost as it is affected by OpenAI server stability.

\noindent
{\bf [Experimental Results]:} 
Table~\ref{tab:program_repair_nus} shows the repair performance of the evaluated three approaches. The x\% (y\%) indicates the results of TOP-5 (AVG-5) (Section~\ref{chatgpt_application}). We note that ChatGPT is a generic and multi-task model. It is not specifically developed for repairing incorrect assignments.
Regarding the TOP-5 metric, APR tool Refactory achieves the highest repair rate of 90\% while ChatGPT achieves competitive results, repairing 84\% incorrect codes and Codex can repair 66\% of them. Specifically, ChatGPT (or ChatGPT\_D) can achieve better results on the second, third and fifth problems than Refactory, with a 100\% repair rate on the problem of Top k Elements. 
Regarding AVG-5, ChatGPT can repair an average of 60\% of incorrect codes which is higher than Codex\_D. 
These results reveal that ChatGPT does not achieve high performance when it is just allowed to generate one patch (once-attempt repair).
On the other hand, when provided with the problem descriptions, Codex\_D can repair more incorrect codes than Codex. However, ChatGPT\_D underperforms ChatGPT except for the second problem, \ie providing a problem description reduces the repair performance of ChatGPT.

\begin{table}[!t]
	\centering
	\caption{Performance of ChatGPT, Codex and Refactory on program repair on incorrect submissions for five common programming assignments from the National University of Singapore.}
	\label{tab:program_repair_nus}
	\resizebox{1\linewidth}{!}
	{
	\begin{threeparttable}
		\begin{tabular}{l|rr|cc|cc|c}
		\toprule
            {\bf Problems} & {\bf \#Lines of Code} & {\bf \#Incorrect} & {\bf ChatGPT} & {\bf ChatGPT\_D} & {\bf Codex} & {\bf Codex\_D} & {\bf Refactory} \\
		\hline
            Sequential Search & 10  & 575 & 82.4\% (58.5\%) & 79.5\% (58.1\%) & 63.7\% (20.2\%) &  71.5\% (24.2\%) & 99.0\%  \\ 
              Unique Dates Months & 28  & 435 & 76.6\% (45.5\%) & \cellcolor{blue!25}80.5\% (46.6\%) & 60.7\% (20.1\%) & 62.5\% (21.3\%) & 78.1\%  \\ 
              Duplicate Elimination & 7  & 308 & \cellcolor{blue!25}98.1\% (82.9\%) & 95.1\% (82.2\%) & 85.4\% (44.0\%) & 89.6\% (47.5\%) & 97.4\%  \\ 
              Sorting Tuples & 9  & 357 & 78.7\% (51.7\%) & 67.5\% (51.4\%) & 60.2\% (27.2\%) & 60.2\% (26.3\%) & 88.2\% \\ 
            Top k Elements & 11  & 108 & \cellcolor{blue!25}100\% (90.4\%) & 99.1\% (88.1\%)  & 60.2\% (21.9\%) & 78.7\% (36.5\%) & 88.0\%  \\ 
            \hline
            Overall & 14  & 1783 & 84.0\% (60.1\%) & 81.2\% (59.9\%) & 
            65.8\% (25.8\%) & 70.6\% (28.7\%) & 90.8\% \\
		\bottomrule
		\end{tabular}
            {$^\ast$ ChatGPT\_D and Codex\_D denote they are provided by the description of the problem.} \\
            {$^\ast$ x\% (y\%) indicates the results of TOP-5 (AVG-5). See randomness in Section~\ref{chatgpt_application}.} 
	\end{threeparttable}
	}
\end{table}
To investigate the cause, we first manually analyze the incorrect submissions that can be fixed by ChatGPT but not by ChatGPT\_D from the remaining four problems. We find that such incorrect code commonly contains bugs that are not governed by the problem descriptions. For example, Figure~\ref{fig:wrong_4_20} shows an incorrect student submission of problem {\em Soring Tuples}. The code is incorrect because it misses the break statement in the second `for' loop which leads to mistakenly repeated elements in the returned list. This output fails to pass the test cases. Meanwhile, the provided problem description does not explicitly exhibit the deduplication requirement. ChatGPT\_D, therefore, is prone to ignore this issue and thus cannot find a way to fix the bug. 
For the incorrect code in problem {\em Unique Dates Months}, we find that some of the functions among them only contain a return statement, \eg unique\_month and contains\_unique\_day as shown in Figure~\ref{fig:wrong_2_213}. For this scenario, providing more information (description) enhances ChatGPT's comprehension and its ability to fix the code, resulting in the superior performance of ChatGPT\_D on this problem. Additionally, ChatGPT\_D struggles to address the exception incurred by empty input, as the provided problem description does not specify the input requirements, drawing ChatGPT's attention to other aspects (non-bug related information).
Overall, ChatGPT has limited attention when addressing a given prompt input. To improve repair performance, we suggest programmers can provide ChatGPT with informative problem descriptions or additional information related to the bug such as failing test cases. To validate our speculation, we manually analyze 35 sampled incorrect codes that cannot be fixed by ChatGPT\_D. By providing problem descriptions and extra failing test cases, 30 (95\% confidence, 5\% error) of them can be fixed for the first time.

\begin{figure}[ht]
    \centering
    \scriptsize
\begin{lstlisting}[language=Python,linewidth={\linewidth},frame=tb,
    basicstyle=\ttfamily,
    basicstyle=\small,
    % baselinestretch=1.2
    ]
(@\bf 1:@) def sort_age(lst):
(@\bf 2:@)       new_lst=[]
(@\bf 3:@)       new_lst.append(lst[0])
(@\bf 4:@)       for i in lst[1:]:
(@\bf 5:@)             for j in range(len(new_lst)):
(@\bf 6:@)                   if i[1]>new_lst[j][1] and j==0:
(@\bf 7:@)                         new_lst.insert(0,i)
(@\bf 8:@)                   elif i[1]<new_lst[j][-1]:
(@\bf 9:@)                         new_lst.insert(-1,i)
(@\bf 10:@)                  elif i[1]>new_lst[j][1]:
(@\bf 11:@)                        new_lst.insert(j,i)
(@\bf 12:@)      return new_lst

\end{lstlisting}
    \caption{An incorrect submission for problem {\em Soring Tuples}.}
    \label{fig:wrong_4_20}
\end{figure}

\begin{figure}[ht]
    \centering
    \scriptsize
\begin{lstlisting}[language=Python,linewidth={\linewidth},frame=tb,
    basicstyle=\ttfamily,
    basicstyle=\small,
    % baselinestretch=1.2
    ]
(@\bf 1:@)   def unique_day(date, possible_birthdays):
(@\bf 2:@)         count=0
(@\bf 3:@)         for i in range(len(possible_birthdays)):
(@\bf 4:@)               if date==possible_birthdays[i][1]:
(@\bf 5:@)                     count=count+1
(@\bf 6:@)         if count>=2:
(@\bf 7:@)               return False
(@\bf 8:@)         return True
(@\bf 9:@)   def unique_month(month, possible_birthdays):
(@\bf 10:@)       return
(@\bf 11:@) def contains_unique_day(month, possible_birthdays):
(@\bf 12:@)       return

\end{lstlisting}
    \caption{An incorrect submission for problem {\em Unique Dates Months}.}
    \label{fig:wrong_2_213}
\end{figure}

\subsection{ChatGPT for Code Summarization}
\label{subsec:rq3}

\noindent
{\bf [Experiment Goal]:} We aim to explore the ability of ChatGPT to explain code. To do so, we evaluate whether ChatGPT can explain the intention of correct and incorrect code against the description of programming problems.

\noindent
{\bf [Experiment Design]:} 
Given a programming problem, programmers aim to achieve the same goal by writing code to satisfy associated specifications (\eg problem description). As a result, the correct and incorrect code they write both reflects the same intention as the problem description. To assess ChatGPT's ability to summarize code, we consider investigating whether ChatGPT can explain the original intention of code even though they contain mistakes or errors.  
To that end, we first request ChatGPT to explain the intention (natural language) of 2443 correct and 1783 incorrect codes in Refactory benchmark respectively (see the prompt in Section~\ref{chatgpt_application}). Second, we employ BERT~\cite{devlin2018bert}, a widely-used pre-trained large language Encoder model known for its capability of understanding and classifying natural language~\cite{khurana2023natural, min2021recent, qiu2020pre, koroteev2021bert}. We use BERT to produce the semantic embedding representations of the above intentions of the code and the associated problem descriptions. Finally, we compute the distribution of similarities between intentions (\ie semantic embeddings) generated for correct code and the ground-truth problem descriptions, which we compare against the distribution of similarities between intentions (\ie semantic embeddings) generated for incorrect code and the ground-truth problem descriptions.
Overall, our objective is to evaluate whether ChatGPT can consistently generate the intention for correct and incorrect code.

\begin{figure}[h]
	\includegraphics[width=.8\columnwidth]{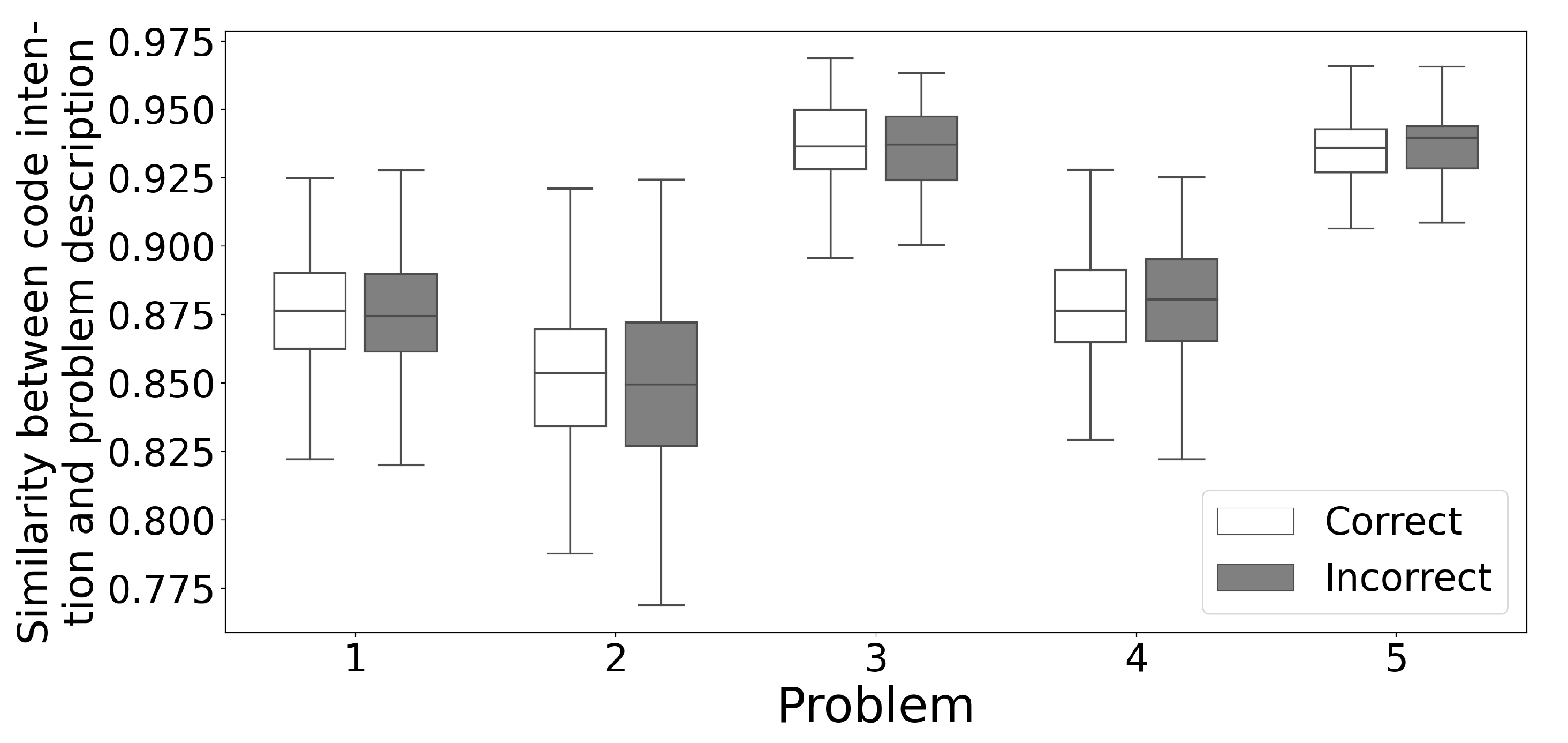}
	\vspace{-3mm}
	\caption{Similarity distribution between intentions of correct code and problem description {\em vs.} similarity distribution between intentions of incorrect code and problem description.}
	\label{fig:distribution1}
\end{figure}

\noindent
{\bf [Experimental Results]:} 
Figure~\ref{fig:distribution1} shows the result of the distribution. The similarity distribution between the intent of incorrect code and the problem description is not significantly different from the similarity between correct code and the problem description. We calculate Mann-Whitney-Wilcoxon (MWW)~\cite{wilcoxon1945individual,mann1947test} and confirm the null hypothesis (no significant difference).

\begin{figure}[h]
	\includegraphics[width=.9\columnwidth]{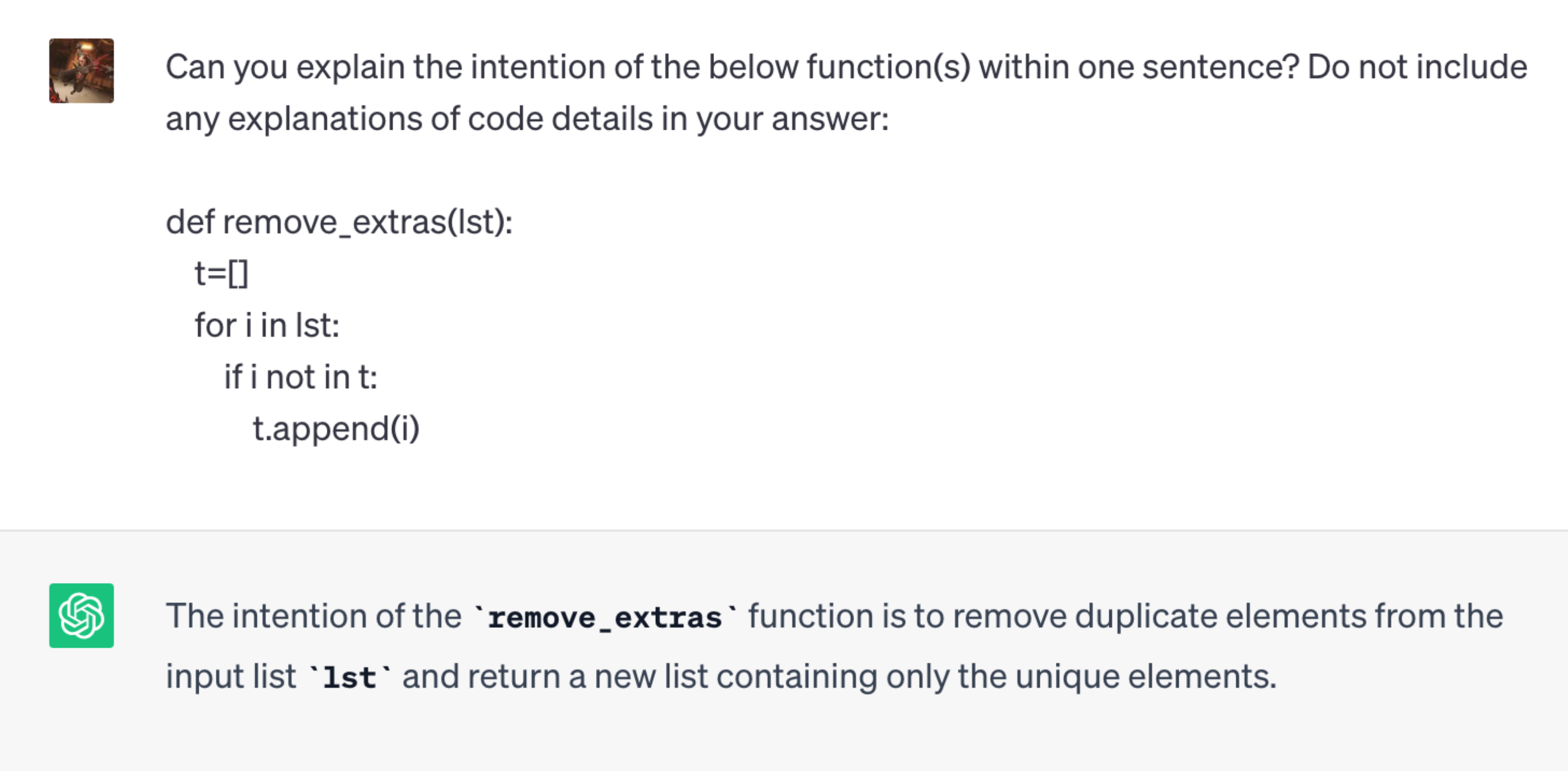}
	\caption{The intention explanation of ChatGPT for an incorrect implementation for the problem {\em Duplicate Elimination}.}
	\label{fig:307}
\end{figure}

Based on the observation, we further explore the characteristics of correct and incorrect code in different problems that allow ChatGPT to successfully infer their intentions. As we see, the similarity distributions in the second problem (Unique Dates Months) are lower compared to others. This discrepancy arises due to the second problem comprising three functions that pose challenges for ChatGPT in summarizing the intention. To investigate the effect of the multi-functions, we separate the three single functions from the code in the second problem to reproduce the experiments.
We find the intentions of (in)correct code can be identified by ChatGPT for functions {\em unique\_day} while not for {\em unique\_month} and {\em contain\_unique\_days}. After manual inspection, we observe that many of the incorrect implementations of functions {\em unique\_month} and {\em contain\_unique\_days} only contain a single return statement (an example in Figure~\ref{fig:wrong_2_213}), increasing the difficulty of intention identification of incorrect code (lowest similarity).
As for the other problems, ChatGPT generates highly similar intentions of incorrect code as correct code and problem descriptions. For instance, the ground truth intention for the third problem is described as {\em ``This function takes in a list and returns a new list with all repeated occurrences of any element removed.''} Figure~\ref{fig:307} shows a typical intention inference example for the third problem (Duplicate Elimination). In this example, the presented code is incorrect due to the lack of a return statement. When being asked about the intention of this code, ChatGPT explains that {\em this function {\bf returns} a new list containing only the unique elements}, even though the return statement is mistakenly missing in this code. To avoid the effect of the function name, we replace the function name with {\em ``a''}. We then still observe the same result.
Overall, the results reveal that ChatGPT is able to identify the original intention of the given (in)correct code. This finding enables ChatGPT to address the oracle problem: in test generation, we do not always have a precise specification of what the output should be. Given a correctness-unknown code snippet without the test suites, ChatGPT can help reason the original intention of the code for programmers. With this intention, programmers can determine the expected output for a test input, and design the test case accordingly. We validate our speculation by analyzing 39 sampled incorrect codes that ChatGPT can identify the defects of 30 of them by reasoning the original intentions and automatically generating test cases. A recent study has demonstrated the potential of ChatGPT to find failure-inducing test cases by reasoning the intention of code~\cite{li2023finding}.

\section{Threats to validity}
\label{sec:threats}

Our empirical study may contain several threats to validity that we have attempted to relieve.

\vspace{0.1cm}
\noindent
{\sc \em Threats to External Validity.} Large language models usually generate varied responses for identical input across multiple requests due to their inherent randomness. One threat to the validity of our study is that conclusions drawn from random results may be misleading. To mitigate this threat, we have requested results from LLMs five times to obtain the TOP-5 and AVG-5 performance metrics and compare them against state-of-the-art approaches. Additionally, we conduct our experiment on a large dataset (\eg 1783 code samples for program repair), which helps reduce the impact of randomness to some extent.
Another potential threat to validity is the selection of benchmarks. We make the choice based on the available large dataset and the need to prevent data leakage in LLMs.

\vspace{0.1cm}
\noindent
{\sc \em Threats to Internal Validity.}
A major threat to internal validity lies in the processing of ChatGPT's responses. Although we request plain code, ChatGPT occasionally generates natural language explanations alongside code in its responses, which can lead to the responses not being able to pass test suites. To mitigate this threat, we employ scripts to remove natural language explanations from the received responses and manually check the code that raises exceptions.
On the other hand, annotations in the original student submissions from Refactory benchmark may impact the evaluation of ChatGPT's code summarization capability on submissions. We also use scripts to remove these annotations from the submissions.
Finally, we release the ChatGPT-generated responses and submissions with annotations removed for the community to review.

\vspace{0.1cm}
\noindent
{\sc \em Threats to Construct Validity.}
In our experiment, our evaluation is based on the GPT-3.5 model. At the time of our submission, the latest model is GPT-4. However, GPT-4 is subject to rigorous request rate restrictions and has garnered reports of instability and inaccuracy. It's worth noting that OpenAI continually updates the behind large language model that supports ChatGPT. Therefore, the results that we have reported in this paper are likely an underestimation of the capability of the ChatGPT models. In future studies, we plan to investigate the GPT-4 version once it establishes itself as a stable research platform.
\section{Conclusion}
\label{sec:conclusion}
In this paper, we performed an empirical study on the ChatGPT generative large-scale language model (LLM), aiming to assess its true potential as an assistant bot for programmers. We investigate ChatGPT on three code-related tasks: code generation, program repair and code summarization. 
In the task of code generation, we evaluate ChatGPT and two prior LLMs on 120 common programming problems sampled from LeetCode 2016-2020 and LeetCode 2022 respectively. The results demonstrate the dominance of ChatGPT while revealing that ChatGPT struggles to generalize to new and unseen problems. We also validate the negative impact of long prompts on the inference capabilities of ChatGPT. 
For the program repair task, we investigate ChatGPT's ability to repair diverse incorrect implementations submitted for common programming assignments from a large-scale benchmark. ChatGPT achieves competitive results compared against Refactory, a state of the art semantic-based assignments repair tool. Additionally, we find that providing prompts that are not related to bug information (\eg generic problem description) makes ChatGPT perform even worse: ChatGPT has a limited attention span. 
For the code summarization task, we evaluate whether ChatGPT can produce consistent intention explanations for (in)correct codes against ground truth problem description. Surprisingly, the results imply that ChatGPT can reason about the original intention of (in)correct code. This insight provides the opportunity to leverage ChatGPT in the testing research field where the test oracle problem remains an open question. 

Overall, we expect the findings and actionable insights of this study on ChatGPT to have a potent impact across the research on software engineering, and in particular for pushing further assistive technology for programmers.

\vspace{0.1cm}
{\bf Data Availability.}
All the artefacts of this study are available in the following public repository: {\bf\url{https://github.com/HaoyeTianCoder/ChatGPT-Study}}



\balance
\bibliographystyle{ACM-Reference-Format}
\bibliography{references}

\end{document}